\documentclass[useAMS,usenatbib]{mn2e}
\usepackage{graphicx}
\usepackage{amsmath}
\usepackage{epstopdf}
\usepackage{epsfig}

\title[Detailed study of the microwave emission of the supernova remnant 3C~396]{Detailed study of the microwave emission of the supernova remnant 3C~396}
\author[A. Cruciani et al.]{A. Cruciani$^{1}$,\thanks{E-mail: angelo.cruciani@roma1.infn.it}
E.S. Battistelli$^{1}$, E. Carretti$^{2,3}$, P. de Bernardis$^{1}$,
\newauthor R. Genova-Santos$^{4,5}$, S. Masi$^{1}$, B. Mason$^{6}$, D. Perera$^{7}$, F. Piacentini$^{1}$, 
\newauthor B. Reach $^{6}$, J.A. Rubino-Martin$^{4,5}$
\\
$^{1}$Department of Physics, Sapienza University of Rome, Piazzale Aldo Moro 5, 00185 Rome, Italy
\\
$^{2}$CSIRO Astronomy and Space Science, PO Box 276, Parkes, NSW 2870, Australia
\\
$^{3}$Cagliari Astronomical Observatory, Via della Scienza 5 - 09047 Selargius, CA, Italy 
\\
$^{4}$Instituto de Astrofisica de Canarias, C/ Via Lactea, s/n, E38205, La Laguna, Tenerife, Spain
\\
$^{5}$Departamento de Astrof\'{\i}sica, Universidad de La Laguna (ULL), 38206 La Laguna, Tenerife, Spain
\\
$^{6}$National Radio Astronomy Observatory, 520 Edgemont Rd, Charlottesville, VA 22903
\\
$^{7}$National Radio Astronomy Observatory, P.O. Box 2, Green Bank, WV}

\begin{document}

\date{Received ...; accepted ...}

\pagerange{\pageref{firstpage}--\pageref{lastpage}} \pubyear{2015}

\maketitle

\label{firstpage}

\begin{abstract}
We have observed the supernova remnant 3C~396 in the microwave region using the Parkes 64-m telescope.
Observations have been made at 8.4~GHz, 13.5~GHz, and 18.6~GHz and in polarisation at 21.5~GHz. We 
have used data from several other observatories, including previously unpublished observations performed by
 the Green Bank Telescope at 31.2~GHz, to investigate the nature of the microwave emission of 3C~396.
Results show a spectral energy distribution dominated by a single component power law emission with $\alpha=(-0.364 \pm 0.017)$. Data do not
 favour the presence of anomalous microwave emission coming from the source.

Polarised emission at 21.5~GHz is consistent with synchrotron-dominated emission. We present microwave 
maps and correlate them with infrared (IR) maps in order to characterise the interplay between 
thermal dust and microwave emission. IR vs. microwave TT plots reveal poor correlation between mid-infrared
and microwave emission from the core of the source. On the other hand, a correlation is detected in the tail emission of the outer shell of 3C~396, which could be ascribed to Galactic contamination.  
\end{abstract}


\section{Introduction}

The nature of the emission of supernova remnants (SNRs) and their interaction with the surrounding medium can be studied with multifrequency observations ranging from the radio to the X-ray regions of the spectrum. In the microwave region, a detailed multifrequency analysis of SNR emission can shed light on the interplay between synchrotron emission, which typically dominates SNR radio emission, and the possible presence of other mechanisms such as anomalous microwave emission (AME, \cite{dra98}). Of key importance is the ability to distinguish between different components within the emitting source as well as to disentangle local and foreground contaminants. To this end, high angular resolution (i.e.\ arcminute-level) observations are crucial and allow one to distinguish different contributions within the same Galactic region (e.g.\ \cite{bat012}).

3C~396 is a shell-like SNR with a mean angular diameter of $7'.8$ \citep{patnaik}. Its distance, estimated 
from \textit{Chandra} data \citep{olbert} and updated by means of CO observations, is about 6.2~kpc, and its age is 
about 3000~yr \citep{su}. \cite{patnaik} performed a complete overview of the numerous radio observations 
of the source before 1990. Using high angular resolution observations at 1.4~GHz obtained with the Very
 Large Array (VLA), \cite{patnaik} were able to identify the presence of two separate components in the 
source: the core, dominated by non-thermal synchrotron emission, and the tail, representing about $10\%$ 
of the total flux at 1.4~GHz with a less steep spectral index compatible with free--free emission. 
\cite{anderson} discussed the spatial spectral index variations of the spectral energy distribution (SED) in this SNR and concluded that these variations do not coincide with features in total intensity, but also found
that the region associated with the brightest feature in the SNR has a somewhat flatter spectral index with respect 
to the average SNR spectral index. The presence of a small synchrotron pulsar wind nebula (PWN) within 3C~396 
has been reported, although there is no spatial correspondence with the radio feature in high-resolution 20-cm VLA 
images of the remnant. This component could [would?] have a contribution of about $4\%$ of the total radio flux density at 
1.4~GHz \citep{olbert}, not enough to explain the spatial spectral index variations of the SED in this SNR.

\cite{sca07} observed 3C~396 with the Very Small Array (VSA) experiment \citep{wat03} and tentatively reported finding 
 anomalously high emission at 33~GHz, which suggested for the first time the possible presence of AME due to 
spinning dust in an SNR. As reported by \cite{pla15}, this evidence could not be confirmed by the \textit{Planck} experiment 
owing to the lack of angular resolution and its vicinity to the Galactic plane. Besides AME, the excess seen by the VSA 
could also  be explained by the presence of a significant level of thermal emission (about $50~\%$ of the total emission at 1.4~GHz),
 causing the flattening of the spectrum at frequencies greater than 10~GHz \citep{onic}. \cite{sca07} emphasize the
 need for further measurements in the 10--20 GHz range, which is among the goals of the present paper.

An attempt to evaluate the amount of free--free emission was performed, using the radio recombination line (RRL) 
survey of the Galactic plane from the H I Parkes All-sky Survey  \citep{alves,alves15}; the presence of 
diffuse free--free emission in the 3C~396 region is quite evident, although the angular resolution is poor
(beamwidth of 14.4~{arcmin} FWHM). A rough estimate of the total flux gives an upper limit of 3~Jy at 1.4~GHz.

Polarisation measurements of the SNR have been made, mainly at 5~GHz, and show the presence of a mean polarisation 
fraction of about $3\%$ \citep{sun}, with some regions being polarised at the 5--10\% level and a small 
outer region being up to $50\%$ polarised \citep{patnaik}.

In this paper we present observations at arcminute-level resolutions of 3C~396 performed with the Parkes
single-dish 64~m telescope at 8.4~GHz, 13.5~GHz, 18.6~GHz, and 21.5~GHz.  \\
In Sections \ref{sec:8.4obs}--\ref{sec:21.5obs} we present the observations performed with the Parkes telescope, 
in Section \ref{gbt} we introduce the previously-unpublished 3C~396 observations performed by the 100-m Green Bank Telescope 
(GBT) and in Section \ref{anc} we briefly introduce the remaining ancillary data used to fit the SED of the source. 
In Section \ref{mor} we study the morphology of the SNR and in Section \ref{sed} we present the SED, concentrating 
on the microwave band. The polarisation data, based on measurements at 21.5~GHz, are presented in Section~\ref{pol}. In Section~\ref{ir} we show the correlation between IR and microwave data.

\section{Observations and data reduction}\label{parkes}

Observations were made in four different frequency bands with the 64-m Parkes Radio Telescope, NSW Australia, 
operated by ATNF-CASS, CSIRO: photometric observations were conducted at three frequencies 
(8.4~GHz, 13.5~GHz, and 18.6~GHz) and polarisation observations at 21.5~GHz. A further unpublished observation 
performed by GBT at 31.2 GHz is presented in the last subsection.

\subsection{8.4~GHz observations}\label{sec:8.4obs}

The 8.4~GHz observations were conducted with the MARS receiver of the Parkes telescope on 26 July 2011 for 
4~h. The receiver is a circular polarisation system with $T_{\rm sys} \sim 30$~K, a resolution of 2.4~arcmin, and a bandwidth of 400~MHz centred at 8.4~GHz. To measure the 
whole useful bandwidth the backend Digital Filter Banks Mark 3 (DFB3) was used with a configuration of 512 frequency channels of 2~MHz resolution for a total bandwidth of 1024~{MHz}. The correlator has full Stokes parameter capability, 
recording the two autocorrelation products $RR^*$, $LL^*$ and the complex cross-product 
 $RL^*$ whose real and imaginary parts are the two Stokes parameters $Q$ and $U$. 
The gain is 1.18~Jy~K$^{-1}$.
The source PKS~B1934-638 was used for flux density scale calibration with an accuracy of 5\% \citep{rey94}.

The channels spanning the 400~MHz band were then binned into twenty 20-MHz sub-bands.  
A standard basket-weaving technique with two orthogonal scan sets along R.A. and Dec.\ spaced by 
45~arcsec  was used to observe an area of $20'\times20'$ centred on the source. The scan speed was 
$0^{\circ}.5$/min, with a sampling time of 0.25~s. For each scan, a linear fit performed off-source is removed. Map-making software based on the \cite{eme88} Fourier 
algorithm was applied to make the maps \citep{carretti10}.  This technique effectively reduces 1/$f$ noise and
removes stripes and features that differ between the two orthogonal sets of scans. 

The twenty sub-band maps were binned together in one map for the analysis. The final
 rms on the map is 22~mJy/beam on a beam-size scale. This is higher than the expected sensitivity 
($\sim$ 0.7~mJy/beam); however, 3C~396 is locate at ($l$,~$b$) = (39$^\circ$.2, -0$^\circ$.3), close to the Galactic plane, and the excess signal could be due to diffuse Galactic emission. To investigate this possibility we analysed the 
1.4~GHz CHIPASS map \citep{cal14}, which has a resolution of 14.5~\textit{arcmin}, and found an rms of 650~mK in the area around 3C~396. 
Following \citep{battistelli2015} we estimate an rms from the Galactic signal at our frequency and resolution of 23~{mJy/beam}, quite consistent with the 
{rms} signal directly measured around the source in our map.

\begin{figure}
\begin{center}
\includegraphics[width=7.5cm]{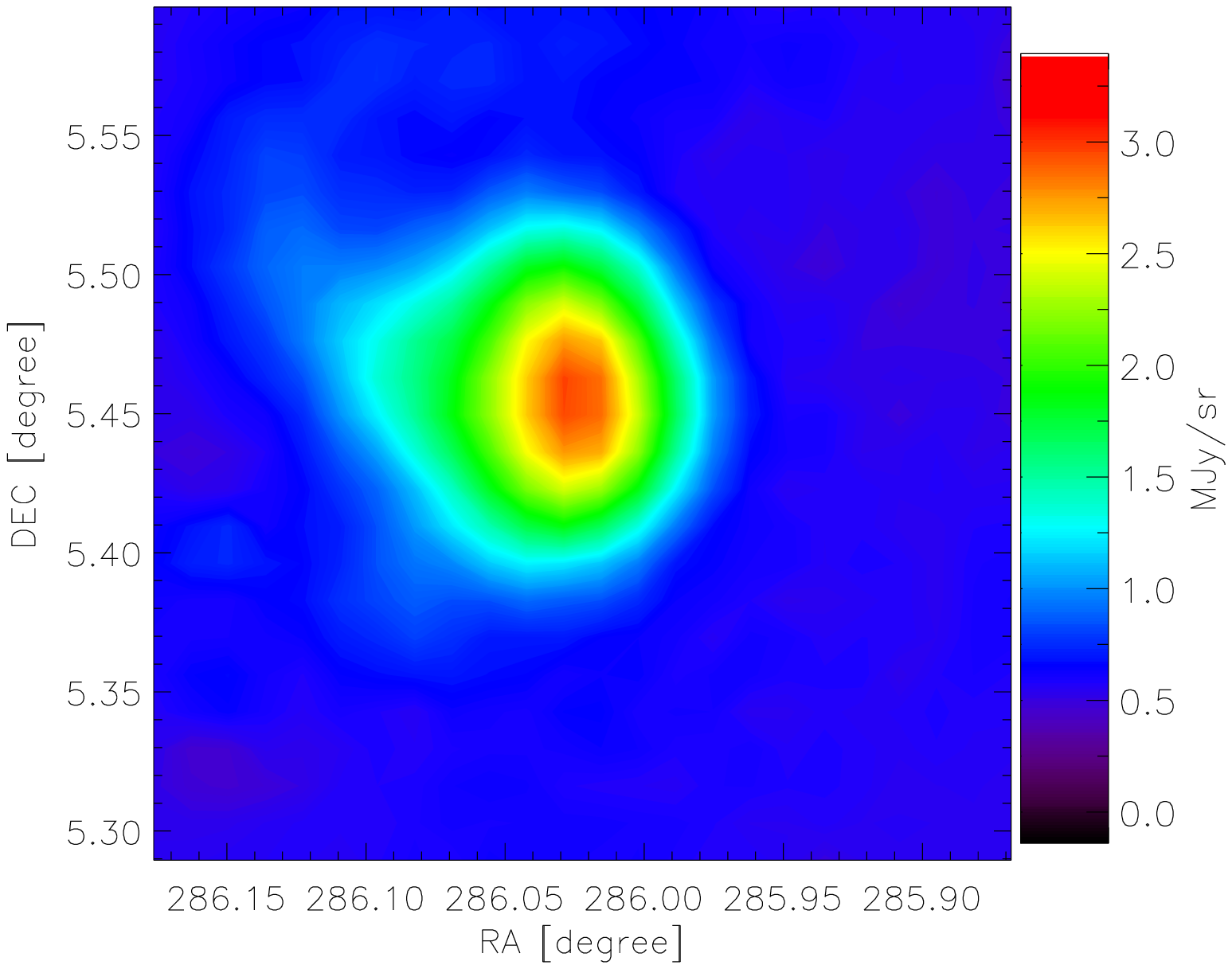}
\caption{8.4~{GHz} map of 3C~396 obtained with the Parkes Radio Telescope using the MARS receiver. The angular resolution is 2.4~arcmin (FWHM) and the final rms of the map is 22~{mJy/beam} (0.040 MJy/sr).} \label{fig1}

\includegraphics[width=7.5cm]{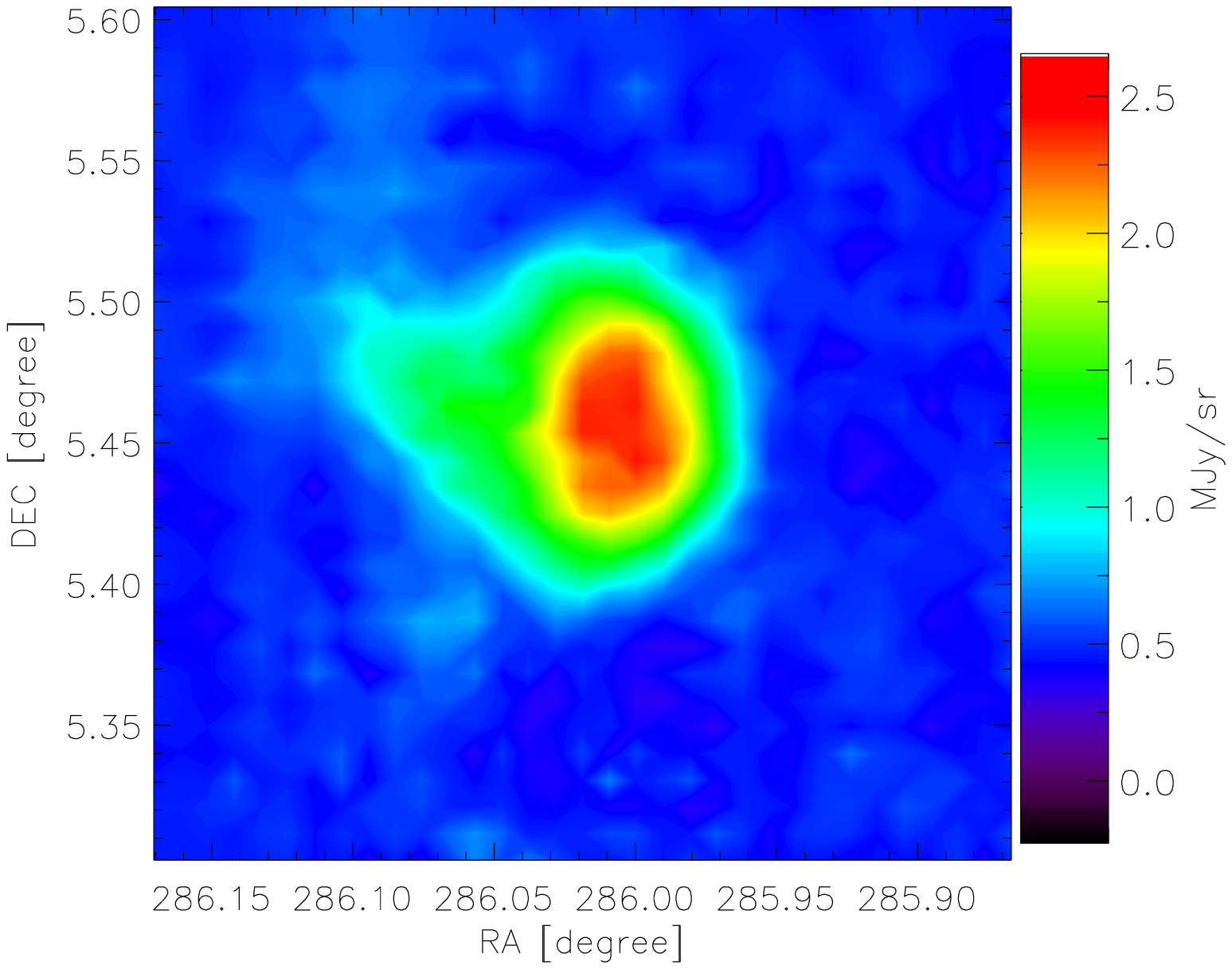}
\caption{13.5~{GHz} map of 3C~396 obtained with the Parkes Radio Telescope using the Ku-band receiver. 
The angular resolution is 1.7~arcmin (FWHM) and the final rms of the map is 18~{mJy/beam} (0.065 MJy/sr).} \label{fig2}

\includegraphics[width=7.5cm]{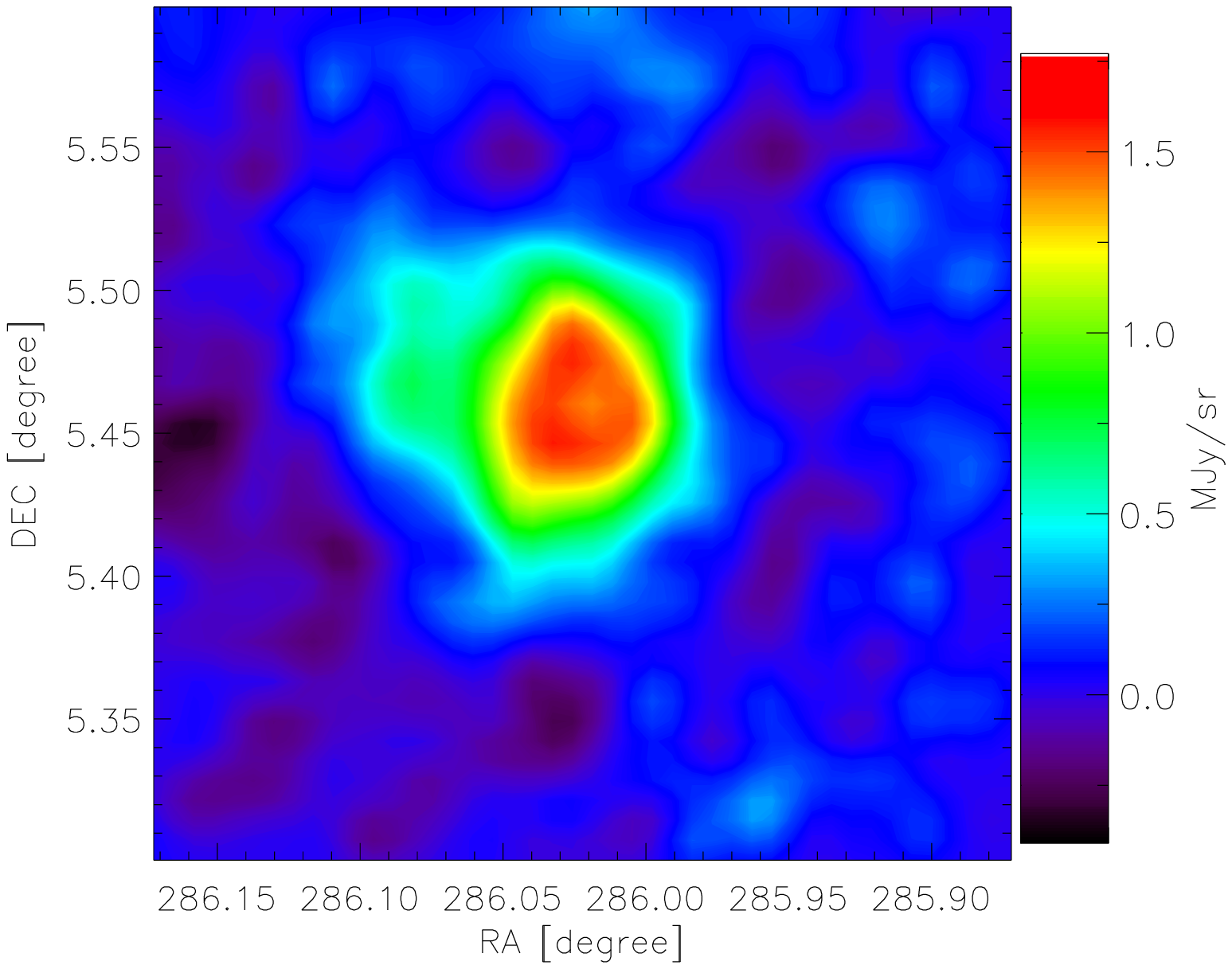}
\caption{18.6~GHz map of 3C~396 obtained with the Parkes Radio Telescope using the 13-mm receiver. 
The angular resolution (after smearing to Ku-band angular resolution) is 1.7~arcmin (FWHM) and the final rms of the map is 32~{mJy/beam} (0.11 MJy/sr).} \label{fig3}

\end{center}
\end{figure}

\subsection{13.5~{GHz} observations}\label{sec:13.5obs}

The 13.5~GHz observations were conducted with the Ku-band receiver of the Parkes telescope on 
31 August 2011 and 1 September 2011, for a total of 3~h. The receiver is a linear polarisation package with 
$T_{\rm sys} \sim 150$~{K}, a resolution of 1.7~arcmin, and a bandwidth of 700~{MHz} 
centred at 13.55~{GHz}. To detect the whole useful bandwidth the backend Digital Filter 
Banks Mark 3 (DFB3) was used with a configuration of 512 frequency channels of 2~MHz resolution for a total bandwidth of 1024~{MHz}. Only the two autocorrelation products were used ($XX^*$, $YY^*$) for Stokes~I measurements.
The gain is 1.55{~Jy~K}$^{-1}$.

The source PKS~B1934-638 was used for flux density scale calibration with an accuracy of 5\% \citep{rey94}.
The atmospheric opacity when observing the calibrator was 0.085~dB at the observing elevation (EL); we correct for this attenuation in all our data. During the observations the opacity ranged from 
0.074 to 0.120~dB (including EL effects) for a maximum variation compared to the constant opacity correction of 0.035~dB (0.8\%), 
with marginal effects on the flux densitiy scale accuracy. 
The opacity at zenith was computed from atmospheric parameters (temperature, pressure, and relative humidity) and used to compute
the opacity at the observing elevation correcting for EL effects (1/cos(EL)). 

The channels spanning the 700~MHz band were binned into seven 100-{MHz} sub-bands for the subsequent map-making
processing. An area of  $20'\times20'$ centred at the source was observed with a standard basket-weaving pattern, in which we performed two orthogonal scan sets along R.A. and Dec. The scans were spaced by 30~{arcsec}, the scan speed was 
0$^\circ$.5/min, and the sampling time 0.25~s. After running the map-making software, the seven sub-band maps were co-added. The final rms signal on the map is 
18~{mJy/beam} on beam-sized scales, larger than the expected sensitivity (4~{mJy/beam}). Following the same 
argument presented in Section 2.1~\ref{sec:8.4obs}, we estimate a Galactic emission contribution of 
{rms}$_{13.5\ \rm GHz}^{1.7'} = $12~{mJy/beam} at the frequency and resolution of our 
observations, consistent with the measured value.

\subsection{18.6~{GHz} observations}\label{sec:18.6obs}

The 18.6 GHz observations were conducted with the 13-mm receiver of the Parkes telescope on 
29 August 2011 for a total of 3~h.
This receiver can be set up either with a linear polarisation feed covering the frequency range 16--26~GHz
or with a circular polariser covering 21.0--22.3~GHz.
We used the linear polarisation configuration with a 800~{MHz} IF band 
centred at 18.6~{GHz}. The system temperature was  $T_{\rm sys} \sim 70$~{K},  
and the resolution was 76~{arcsec}. The backend DFB3 was used 
with with a configuration of 512 frequency channels of 2~MHz resolution for a total bandwidth of 1024~{MHz}. 
Only the two autocorrelation products were used ($XX^*$, $YY^*$) for Stokes~$I$ measurements. 
The gain is 1.56{~Jy~K}$^{-1}$.

The source PKS~B1921-293 was used to calibrate the flux density scale with an assumed flux of 17.2~Jy,
and accuracy of 10\%. This is a variable source on a time scale of a few weeks and its flux density was measured 
with the Australia Telescope Compact Array the day after the Parkes observations (31 August 2011).  
The opacity when observing the calibrator was 0.19~dB at the observing elevation, and we corrected all our data for this attentuation. During the observations the opacity ranged from 
0.18 to 0.28~dB (including EL effects) for a maximum variation compared to the constant opacity correction of 0.09~dB (2\%),
with marginal effects on the flux density scale accuracy.

The frequency channels spanning the 800~MHz band were binned into four 200-MHz sub-bands for the subsequent map-making
step. An area of  $20'\times20'$ centred at the source was observed with a standard basket-weaving pattern, in which we performed two orthogonal scan sets along R.A. and Dec. 

The scans were spaced by 24~{arcsec}, the scan speed was 
0$^\circ$.5/min, and the sampling time 0.25~s. The map-making software 
was run and the four sub-band maps were co-added into a single map. 
We reached an rms sensitivity of 32~mJy/beam on beam-sized scales, after smearing the map to the angular resolution to the Ku-band observations,
larger than the expected sensitivity (2~{mJy/beam}). As described in 
Section~\ref{sec:8.4obs}, we estimate a Galactic emission contribution of 
{rms}$_{18.6\ \rm GHz}^{1.7'} = $12~{mJy/beam} at the frequency and final resolution of our 
map. This can only partly explain the rms signal excess we measure. Possible sources of the additional noise are atmospheric emission fluctuations or 
a 1/$f$-noise contribution from the receiver amplifiers. We do not have enough data to determine which
 of the two effects might have the greater contribution.

\subsection{21.5 GHz observations}\label{sec:21.5obs}

The 21.5~{GHz} observations were conducted with the 13~{mm} receiver of the Parkes telescope on
30 August 2011  for a total of 3~h. We used the circular polarisation set-up with a  900~{MHz} band
centred at  21.55~{GHz}. The system temperature was  $T_{\rm sys} \sim 95$~{K},  
and the resolution was 67~{arcsec}. The backend DFB3 was used 
with a configuration of 512 frequency channels of 2~MHz resolution for a total bandwidth of 1024~{MHz}.
All autocorrelation and complex cross-products of the two circular polarisations were recorded ($RR^*$, $LL^*$, $RL^*$). 
The gain was 1.70~{Jy}~{K}$^{-1}$. Observations consisted of 111 repeated scans of one strip through the source 
at fixed DEC=5$^{\circ}$.43 from RA=285$^{\circ}$.90 to RA=286$^{\circ}$.20 and back. 

The source PKS~B1921-293 was used to calibrate the flux density scale with an assumed flux of 16.5~{Jy},
and accuracy of 10\%.   
The opacity when observing the calibrator was 0.47~dB at the observing elevation, and all our data are corrected for this attenuation. During the observations the opacity ranged from 
0.39 to 0.65~dB (including EL effects) for a maximum variation compared to the constant opacity correction of 0.19~dB (4\%). 
Combined with the flux scale accuracy (10\%) this gives a final accuracy of 11\%.

Off-axis instrumental polarisation can be as high as $0.6\%$ in single observations. Cancellation effects
 \cite[e.g., ][]{car04,ode07} and the rotation of the parallactic angle during the execution of the scans
 reduces this effect when averaging scans taken at different parallactic angles. Following Battistelli et al (2015), 
we estimated a level of $0.2\%$ after averaging all scans. Combining with the on-axis term, the overall
 systematic residual instrumental polarisation is estimated at $0.3\%$.

The frequency channels over the 900~{MHz} band were binned into 90 sub-bands for flux and instrumental 
polarisation calibration. All the sub-bands were then combined for the subsequent analysis. 
We reached a sensitivity per beam-sized pixel of $\sigma_{Q,U}^{21.5\ \rm{GHz}}$~=~0.2 {mJy/beam} in polarisation, 
consistent with the expected value. The fluctuations in Stokes $I$ are larger with an {rms} of 6.0~{mJy/beam},
consistent with the {rms} of the Galactic signal that, following the procedure of Section~\ref{sec:8.4obs}, 
is estimated to be {rms}$_{21.5\ GHz}^{67"'} = $5.2~{mJy/beam} at the frequency and resolution of 
the observations. The longer exposure time allowed the 1/$f$ and atmospheric noise terms to be reduced below the Galactic diffuse emission confusion,
making the latter the leading term of the noise budget.

\subsection{Ka-band - Green Bank Telescope}\label{gbt}

\begin{figure}
\begin{center}
\includegraphics[width=9cm]{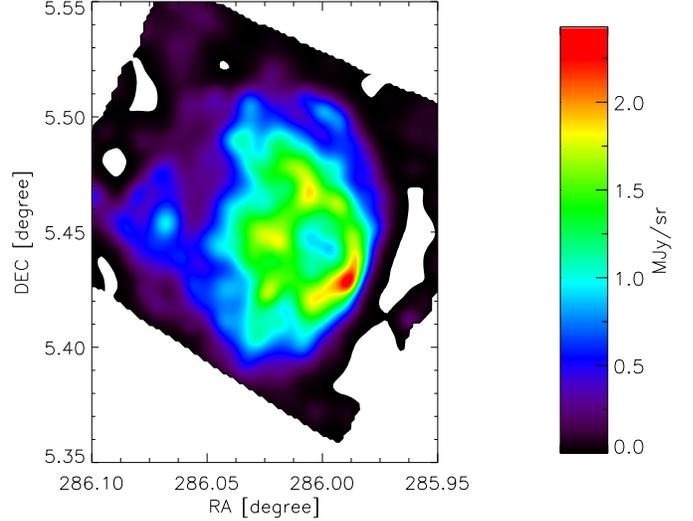}
\caption{31.2~GHz map of the core of 3C~396 obtained with the GBT Ka-band receiver. 
The angular resolution is 0.4~{arcmin} (FWHM).} \label{fig4}
\end{center}
\end{figure}

A $\sim 9' \times 9'$ map of 3C~396 was made at 31.2~GHz under project AGBT10A-001 using the GBT Ka-band receiver
 \citep{jew04} and Caltech Continuum Backend (CCB).\footnote{http://www.astro.caltech.edu/~tjp/GBT/ .} The Ka-band
 receiver is a beam-switched pseudo-correlation receiver which, when used with the CCB, provides excellent continuum 
sensitivity; technical details of these systems are given in \cite{mason09}.  The GBT beam at 31~GHz is $24$~{arcsec} 
(FWHM), and the two (differenced) beams read out by the CCB are separated by $78$~arcsec in the cross-elevation 
direction on the sky. Data are calibrated using 3C~286, referenced to the \textit{WMAP} absolute measurement of
 Jupiter \citep{wei11} There is a $15\%$ calibration uncertainty, dominated by uncertainties in the overall beam area of the GBT. 
The sky map (see Figure~\ref{fig4}) was estimated from the beam-switched data using a maximum entropy method (MEM) 
deconvolution (e.g., \cite{cor88} and \cite{nen86}).

\section{Ancillary data}\label{anc}

Radio emission from 3C~396 has been investigated in depth in the past. We selected a number of previous observations,
 taking into consideration the quality, accuracy and angular resolution of the data. In Table~\ref{tbl-2} we show a summary 
of the data we used. When multiple observations are available, the flux at a given frequency is the weighted average of all the different measurements.

\cite{patnaik} compiled a detailed list and analysis of the observations made before 1990. We include 
all these data. We were also able to retrieve the	 original map performed with the VLA at 1.468~GHz (with a beam FWHM of
 54.4~arcsec and a noise rms of 16~mJy/beam) and at 4.860~GHz (with a beam FWHM of 16~arcsec and a noise rms of 2~mJy/beam),
 which will be used in the morphology section. In addition, we include more recent observations made with the Effelsberg telescope 
\citep{rei97} as in \cite{scaife}. Recently, the SNR was observed at 5~GHz (in intensity and polarisation) by the Sino--German 
survey with the Urumqi telescope \citep{sun}. The results are consistent with other observations in the literature at the same 
frequency. We analysed further observations at 8, 10, and 14~GHz (\cite{langston}, \cite{handa}), which, however, were excluded owing to the low S/N ratio.

We also include the 33~GHz measurement presented in \cite{sca07}. Any comparison of data from Parkes and the Very Small Array (VSA) interferometer needs to take into account the different experiment architecture and their different sensitivities as a function of angular scale. 
A linear fit was subtracted from each linear scan in the Parkes observations to reduce atmospheric contamination and 1/f detector noise. Due to this procedure these maps are sensitive to angular scales roughly between the beam size and the map size (20~arcmin). The VSA observation was made in the extended configuration with a primary beam FWHM of $\approx 72$~arcmin and a synthesized beam FWHM of $\approx 6$~arcmin. (More details of the experimental set-up and data reduction can be found in \cite{dickinson04}.)
The VSA observation is more sensitive than the Parkes observation for scales between 20 and 72~arcmin, though Parkes has a finer angular resolution.
In order to compare the same angular scales of in our two observations, each of which has a different window function, we reanalysed the VSA data. First, we filtered the VSA visibilities using the Parkes window function. Then, using standard
 \textsc{aips} \citep{greisen94} routines, we produced a sky map which we cleaned down to about $1.5$ times the map 
noise sensitivity. In order to derive the flux densities of 3C~396 and the nearby HII region NRAO 591 simultaneousely, we 
applied a double Gaussian fit. The resulting 3C~396 flux density is $5.20\pm 0.33$~Jy. This value is lower than that 
derived in Scaife et al. (2007), $6.64\pm 0.33$~Jy, as a consequence of the filtering process. 

We also included in our analysis \textit{Planck}-HFI observations between 100 and 853~GHz and \textit{Herschel Space Observatory} 
data between 600 and 4300~GHz in order to properly characterise the contribution from thermal dust in the 
microwave region. \textit{Planck}-LFI data of 3C~396 are not included owing to the lack of angular resolution and the 
vicinity to the Galactic plane \citep{pla15}.

\textit{Planck} DR2 maps \citep{cpp1}, downloaded from the Planck Legacy 
Archive,\footnote{http://pla.esac.esa.int/pla/} were used to derive fluxes in the six frequency bands (100, 143,
 217, 353, 545, and 857~GHz) of the High Frequency Instrument (HFI). These maps are supplied in the HEALPix
 pixelization \citep{gorski05} at a resolution of $N_{\rm side}=2048$. The fluxes shown in Table~1 
were obtained within an aperture of radius 0$^\circ$.14 centred at
 $(l,b)=(39^{\circ}.22,-0^{\circ}.31)$, after subtracting a median background level calculated in an external 
annulus between radii $0^{\circ}.14$ and $0^{\circ}.18$. Owing to the proximity of the Galactic plane, the 
uncertainties of these fluxes are dominated by background fluctuations rather than by instrumental noise. 
The flux uncertainties are estimated from the background fluctuations in the ring, taking into account the number of independent samples and the number of pixels in the aperture.

3C~396 was also observed by the \textit{Herschel Space Observatory} during the project Herschel infrared Galactic Plane Survey
(Hi-GAL) \citep{mol10}. We derive flux in five bands  (600, 857, 1200, 1870, and 4300~GHz), processing the data as in \cite{tra11}. The fluxes were calculated using aperture photometry in the same manner as the Planck HFI case described above. In order to calculate the uncertainty, which is dominated by background fluctuations, we split the annulus 
used to subtract the background in ten radial sub-rings and calculated the dispersion of the recovered fluxes using this ensemble of sub-rings. 
\textit{Planck} and \textit{Herschel} flux estimates are strongly affected by the foreground emission evident in their maps. This
 makes them only tentative detections and could be interpreted as upper limits of the mm and sub-mm emission from 3C~396 SNR. 

\begin{table}
\caption{Integrated fluxes used for the 3C~396 SED fit.}\label{tbl-2}
\begin{center}
\begin{tabular}{lcll}
\hline\hline
Frequency  &  Flux &  Reference \\
(GHz)         &  (Jy)  &  \\
\hline  
0.160 &  $35.9 \pm 4.3 $ &  (a), (b) \\
0.408 &  $27.0 \pm 3.0 $ &  (c), (d) \\
0.750 & $18.2 \pm 1.8$   &  (e)\\
1.40   & $14.9 \pm 1.3$   & (e), (f), (g), (h), (i), (l)\\
1.70   & $14.5 \pm 0.8$   & (m)\\
2.70   & $10.9 \pm 0.5$   & (g), (l), (n), (o), (p), (q), (r)\\
3.24   & $11.4 \pm 0.7$   & (s) \\
4.87   & $8.5 \pm 0.9$     & (t)\\
5.00   & $8.84 \pm 0.53$ & (e), (g), (j), (k), (q), (u), (v)\\
6.63   & $10.2 \pm 0.7$   & (s)\\
8.40   & $8.60 \pm 0.43$ & This work \\
10.6   & $6.32 \pm 0.8$   & (s)\\
13.5   & $6.30 \pm 0.31$ & This work \\
18.6   & $4.0 \pm 1.2$ & This work \\
33.0   & $6.64 (5.20) \pm 0.33$ & This work (reviewed)\\
100    & $1.0 \pm 3.1$     & This work \\
143    & $1.7 \pm 3.7$     & This work \\
217    & $10 \pm 14$       & This work \\
353    & $53 \pm 61$     & This work\\
545    & $230 \pm 210$   & This work\\
600    & $280 \pm 150$   & This work\\
857    & $750 \pm 650$ & This work \\
857    & $740 \pm 380$ & This work\\
1200   & $1890 \pm 780$& This work\\
1874   & $2900 \pm 1200$& This work\\
4293   & $1500 \pm 210$& This work\\
			\hline
			\end{tabular}
\begin{tabular}{p{8 cm}}
\tiny
(a)\cite{dul75} (b)\cite{sle77} (c)\cite{sha70} 
(d)\cite{fan74} (e)\cite{kel69} (f)\cite{mil79}
(g)\cite{alt70} (h)\cite{sha76} (i)\cite{patnaik}
(j)\cite{rei70} (k)\cite{sun} (l)\cite{rei97} (m)\cite{dow81} (n)\cite{mil69} 
(o)\cite{hor69} (p)\cite{rei84} (q)\cite{gar75} 
(r)\cite{day70} (s)\cite{hug69} (t)\cite{alt78}
(u)\cite{bin81} (v)\cite{mil75} (w) \cite{scaife}
\end{tabular}
\end{center}
\end{table}

\section{Morphology}\label{mor}

3C~396 has a complex morphology that makes it difficult to determine the nature of its observed emission.
The SNR is a shell-type object with highly non-uniform emission, and
\cite{patnaik} observed the presence of a tail at $RA \simeq 286^{\circ}.07$, $DEC \simeq 5^{\circ}.47$
that accounts for about $10\%$ of total integrated flux at 1.4~GHz. The tail emanates from the eastern 
side of the remnant and  curves more than 120$^{\circ}$ before falling below detectability. Parkes measurements 
have been able -- for the first time in the microwave -- to partially resolve the structure of the SNR and its
 tail. A rough estimate of its contribution at the observed frequencies (8.4, 13.5 and 18.6~GHz) yields 10--20\% of the total integrated flux. Using a TT-plot with the VLA 1.4 GHz VLA map and Parkes 13.5 GHz Parkes map and together with the definition of tail shown in Fig.~\ref{regioni}, we find a spectral index for the tail of $(-0.16 \pm 0.02 (stat) \pm 0.05 (cal))$. This is compatible with free-free emission, as previuosly suggested by \cite{patnaik}.

\begin{figure}
\begin{center}
\includegraphics[width=7cm]{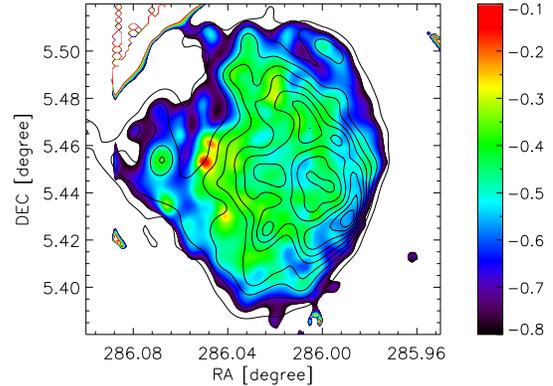}
\caption{Spectral index map of 3C~396 calculated using the 4.8~GHz VLA map (beam FWHM of  16~arcsec) 
and the 31.2~GHz GBT map (beam FWHM of 25~arcsec). Contour levels are from the 31.2~{GHz} map and are at  0.23 , 0.46, 0.68, 0.91, 1.14, 1.37, 1.59, 1.82, 2.05 and 2.28 MJy/sr.} \label{simap}
\end{center}
\end{figure}

In order to investigate the nature and the spatial properties of the emission, we made a spectral
 index map (Figure \ref{simap}) from the two maps with highest angular resolution (VLA at 4.8~GHZ 
and GBT at 31.2~GHz). The two maps where degraded to the same angular resolution and the spectral index
 was derived pixel-by-pixel using the relation $f(\nu)=k\nu^{-\alpha}$. The data at both frequencies
 have a high S/N ratio, and so the error in the fitting procedure is dominated  by systematic errors 
(e.g., calibration error). This map is useful for investigating variations in the spectral index 
within the source itself: it exhibits a steeper spectral index ($-0.6< \alpha <-0.4$) 
in the outer part of the core than in the inner part, where $\alpha$ is between $-0.4$ and $-0.3$. This confirms claims  by \cite{anderson} from  data at 1.4 and 5~GHz.

The observation of such spectral variations in shell remnants can be associated with spatially-dependent particle 
acceleration or bends in the relativistic electron energy spectrum. We cannot exclude the possibility, however, that these 
variations are caused by free--free presence within the source, as suggested by \cite{onic}, or along the line of sight.
 
Analysis of the IR and sub-mm maps (see Fig.~\ref{fig:tt_plots}) retrieved from the \textit{Herschel} and \textit{SPITZER} experiments and ranging from
 500~$\mu$m to 24~$\mu$m shows a lack of emission towards the core of 3C~396, while there is evidence of emission
 in the surrounding area, especially around the tail where the emission peaks at around $160$~$\mu$m. It should be 
stressed, however, that the closeness of the source to the Galactic plane makes its emission at all wavelengths 
fairly contaminated by the Galactic plane itself. It is therefore difficult to disentangle local and diffuse emission. 
A cross-correlation analysis between IR and microwave emission is performed in section \ref{ir}.
 
\section{Spectral Energy Distribution}\label{sed}

In order to retrieve information about the overall emission of 3C~396 we have studied the SED of the SNR. To compare fluxes from all our available data, we reduced the maps in the same way: aperture photometry was performed with a radius of 8.5~arcmin after subtracting a background measured in an annulus with inner radius of 8.5~arcmin and outer radius of 10~arcmin. Reported errors are the combination of calibration error and map noise, estimated from background fluctuations.

\begin{figure}
\begin{center}
\includegraphics[width=8cm]{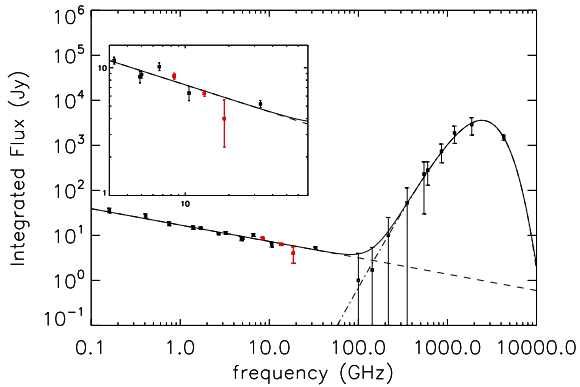}
\caption{SED of 3C~396 including Parkes (in red) and the reprocessed VSA flux at 33~GHz. The fit 
indicates that one synchrotron component is the best explaination of the SED behaviour in the microwave range.
 The fit does not favour the presence of any spinning dust component or further power law components (e.g., free--free).} \label{fig_sed}
\end{center}
\end{figure}

Our baseline model of the SED has one power law component, $Sy$, and a thermal dust emission. Millimetre, sub-millimetre and FIR emission is clearly dominated by thermal dust emission, which we modelled as modified blackbody emission with fixed dust emissivity spectral index $\beta=1.6$ \citep{plan} and free parameters for the temperature $T_{\rm d}$ and the optical depth at 100 $\mu$m, $\tau_{100}$.
The resulting model, which we fit with a routine based on the IDL MPFIT one \cite{mar09}, is therefore: 
\begin{equation}
S(\nu)= Sy \cdot \nu^{\alpha_{1}}+ \tau_{100} \left( \frac{\nu}{3 \rm THz} \right)^{\beta}\cdot BB_{\nu}(T_{\rm d}) \Omega
\end{equation}
where $BB_{\nu}$ is the Blackbody brightness and $\Omega$ is the total solid angle of the source.
Results are shown in Figure~\ref{fig_sed}. The SED of 3C~396 is well-described in the 1-30~GHz range Crab-like synchrotron component with $\alpha=(-0.364 \pm 0.017)$. The dust temperature is $(25.1 \pm 1.2)$~K and  $\tau_{100}$ is $(3.6 \pm 1.2) 10^{-4}$. The fit has $\chi$-square = 24.6 (DOF=21).

We also tested more complex models by adding AME, using a fiducial model derived from \cite{dra98}, or including contaminations from free-free or a second power law component. To select the best model, we used the Akaike information criterion (AIC, \cite{akaike, akaike2}), in which AIC=$2k-\chi^{2}$, where k is the number of estimated parameters. These three modified models (AIC $\approx$ 34.5) are all rejected by AIC with respect to the baseline model (AIC $\approx$ 32.5).

We used the high angular resolution maps of VLA, Parkes and GBT to make a further SED fit restricted only to the core of 3C~396. Its flux was estimated from aperture photometry with a radius of 4.5~arcmin and background subtraction in an annulus with inner radius of 4.5~arcmin and outer radius of 6~arcmin (excluding the tail region). We obtain $\alpha =(-0.466 \pm 0.024)$ and find no anomalous emission.

Infrared and mm-wave measurements can constrain the contribution from dust at the frequencies where AME should be dominant (i.e., below 100~GHz). However, tt is important to stress that 3C~396 is not unambiguously detected in these maps. Consequently the estimated fluxes can only be taken as upper limits on the SNR emission.

\section{Polarisation} \label{pol}

\begin{figure}
\begin{center}
\includegraphics[width=8cm]{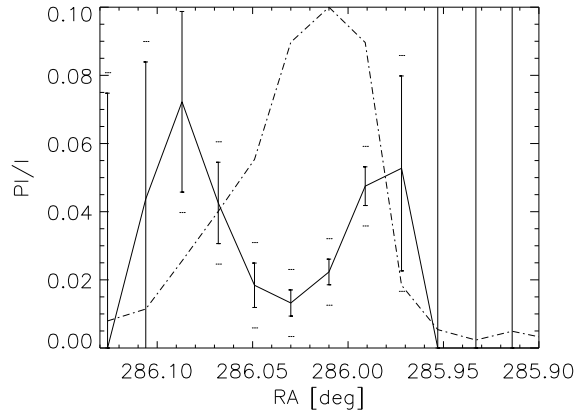}
\caption{Observation of the fraction of polarised intensity over the total intensity at 21.5~GHz 
(fixed declination DEC=$5^{\circ}.43  $), observed with Parkes. Error bars are quoted at the 2~$\sigma$
 level. The dot--dashed line is the profile of the total intensity emission of the source at 21.5~GHz in arbitrary units.} \label{fig_pol}
\end{center}
\end{figure}

Polarisation-sensitive observations at the fixed declination of $5 ^{\circ} 26' 6''$ were conducted at 21.5~GHz with the 13~{mm} receiver of the Parkes telescope. In Figure~\ref{fig_pol} 
we show the percentage polarised emission obtained from the combination of our one-dimensional scans. 
Only the polarisation fraction is presented since the measurements are affected by gain fluctuations on 
long time scales ($>$ 6 hrs) and cannot be reliably calibrated. As previously described, 
spurious polarisation and depolarisation are verified to be less than $0.3\%$. Any estimate of 
polarised intensity ($PI$), and consequently of polarisation fraction, is intrinsically overestimated 
unless a debasing procedure is applied (\cite{war74}). For this reason, we adopted a debiasing
 procedure described in \cite{battistelli2015}, which is based on the Bayesian approach of \cite{vai06}. For
 high signal-to-noise data ($PI/\sigma > 5$), we set a debiased polarised intensity of 
$\sqrt{(PI^{2}-\sigma^{2})}$ as suggested by \cite{war74}. Some of our polarised measurements are 
in the low signal-to-noise regime in which the debiasing is particularly complex. For $PI / \sigma < 2$ 
we set upper limits, while in the intermediate range we integrate the posterior probability density 
function over the parameter space of the true polarisation.

We detect a polarisation of about $5~\%$ in the outer regions of the source, while in the core values 
are lower ($1-4~\%$). These results are in good agreement with previous measurements at
 5~GHz (\cite{patnaik}, \cite{sun}). \cite{sun} observed a mean polarisation in the same one-dimensional 
scan of the source of $(2.4 \pm 0.4) \%$, which is compatible with our average measurement of 
$(2.4 \pm 0.1$(rand)$\pm 0.3$(sys))$~\%$. This result seems to discount the possibility of a large amount 
of free--free emission at 1~GHz (of about $40 \%$) as suggested by \cite{onic}, since one that would require the polarisation at 21.5~GHz to be about half of that at 5~GHz. Further, we can assume that the nature of 3C~396 emission does not dramatically change from radio to microwave frequencies.

\section{Infrared Correlation}\label{ir}

Studying the possible spatial correlations between our microwave data and 
IR data is of interest since the correlation between AME and thermal dust emission is well known  (e.g. \cite{dav06},
 \cite{vid07}, \cite{dic07}, \cite{tod10}, \cite{gen11}, \cite{tib12}, \cite{dic13}). In the case of 3C~396 any correlation between our microwave data and 
IR data would hint at the presence of AME below the detection limit.\

The region emissivity is usually defined as the ratio between the source emission at a frequency near the spinning dust peak ($\sim$20--30~GHz) and the FIR signals at 100~$\mu$m. 
\cite{dic13} reports HII region emissivity in units of $\mu K/(MJy/sr)$ for microwave 
emission in the range 10 to 70~GHz relative to IRAS 100~$\mu$m. These show emissivities 
of the same order of magnitude as high latitude diffuse cirrus (i.e.,\ $\sim10 \mu$K/(MJy/sr)), though it is significantly lower in some regions. Pixel-by-pixel correlations measured by \cite{vid07} 
in $\rho$ Ophiuchi and LDN1780 show a clear correlation with a trend favouring the short 
wavelengths IR bands (i.e.,\ 60~$\mu$m and 25~$\mu$m). Recently, the \textit{Planck} experiment studied all-sky 
correlations, showing an excellent correlation (Pearson $r=0.98$) with the map at 545~GHz and an 
average emissivity of about 70~$\mu$K/(MJy/sr) \citep{pla_ame}. The analysis of selected known 
AME sources has significant correlation (Pearson $r>0.6$) and similar values of emissivity.

\begin{figure}
\begin{center}
\includegraphics[width=8cm]{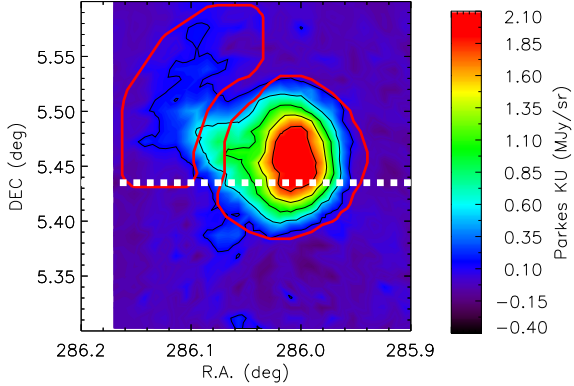} 
\caption{Parkes 13.5~{GHz} map of 3C~396 with superimposed indications of our
  region selection and the polarisation scan path. In order to perform a
meaningful correlation analysis and to obtain the TT-plots shown in
Figure~\ref{tt}, we have divided 3C~396 source into two regions, one
covering the core (central elliptical contour) and one covering the tail
(lateral semicircular contour). The white dashed line illustrates the
path of the polarisation scan reported in Figure~\ref{fig_pol}. Contour levels, from the 13.5~{GHz} map, are at 0.2, 0.70, 1.2 , 1.70, and 2.2 MJy/sr.}
\label{regioni}
\end{center}
\end{figure}

\begin{figure*}
\begin{center}
\includegraphics[width=7cm]{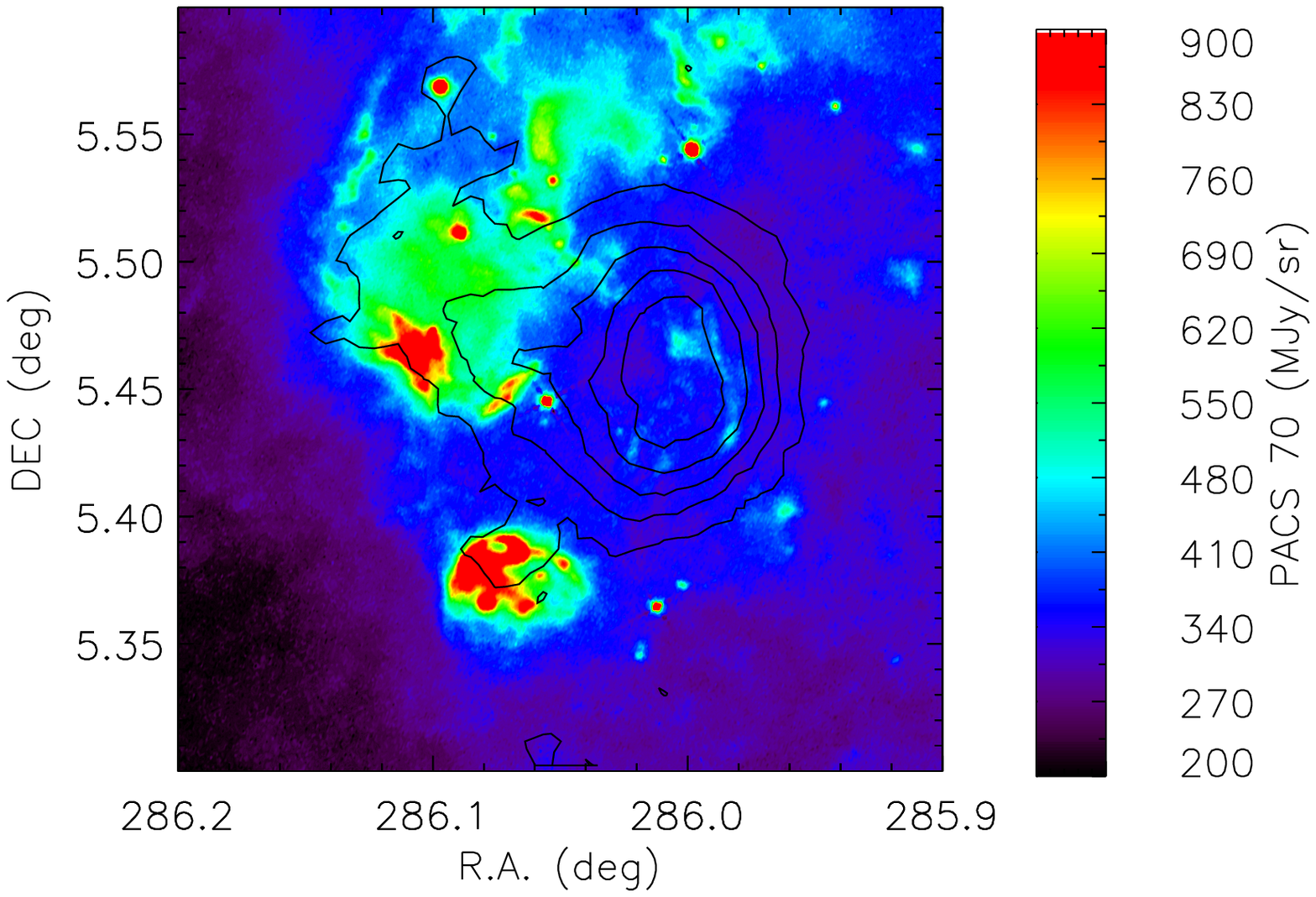}
\includegraphics[width=7cm]{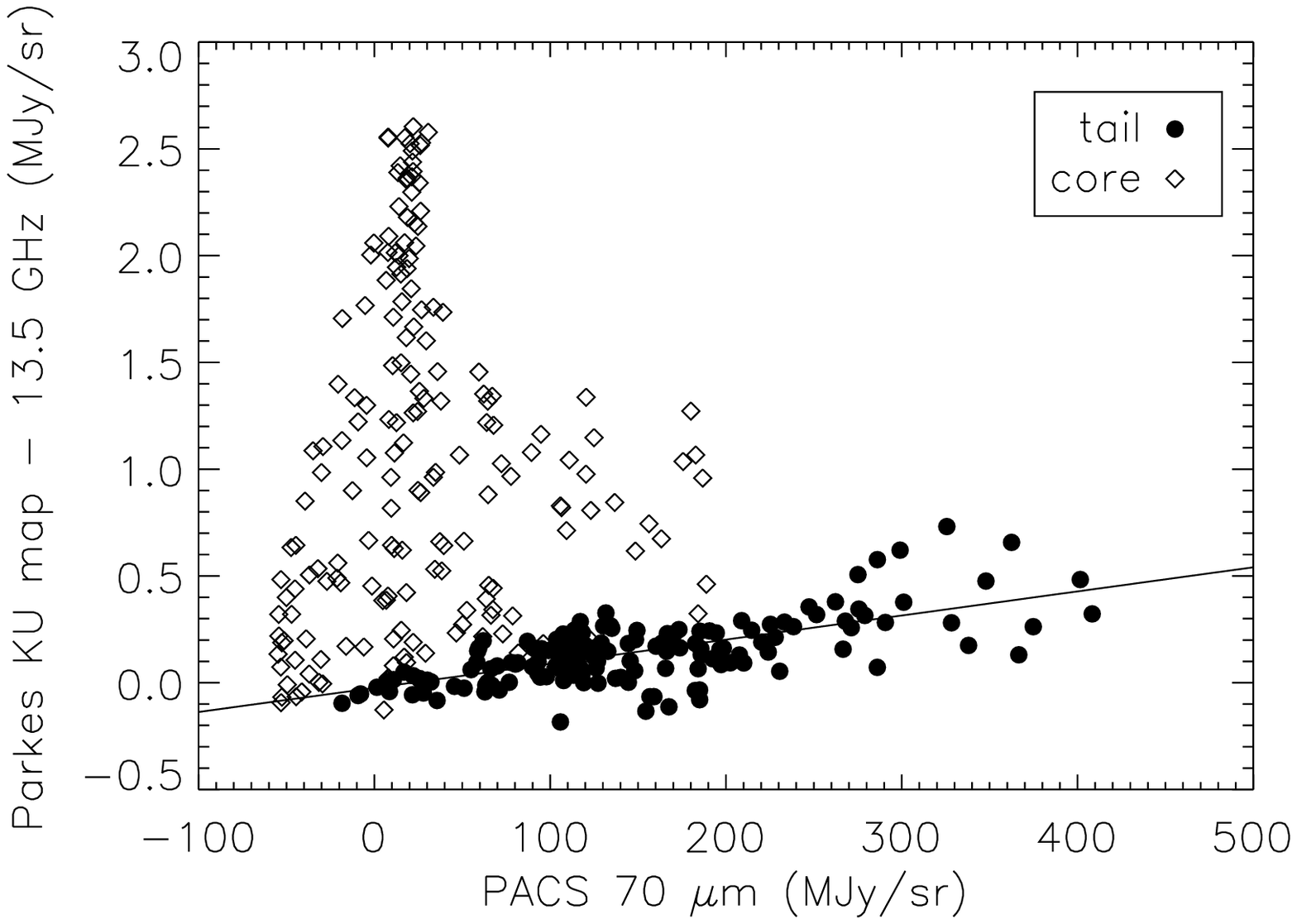}

\includegraphics[width=7cm]{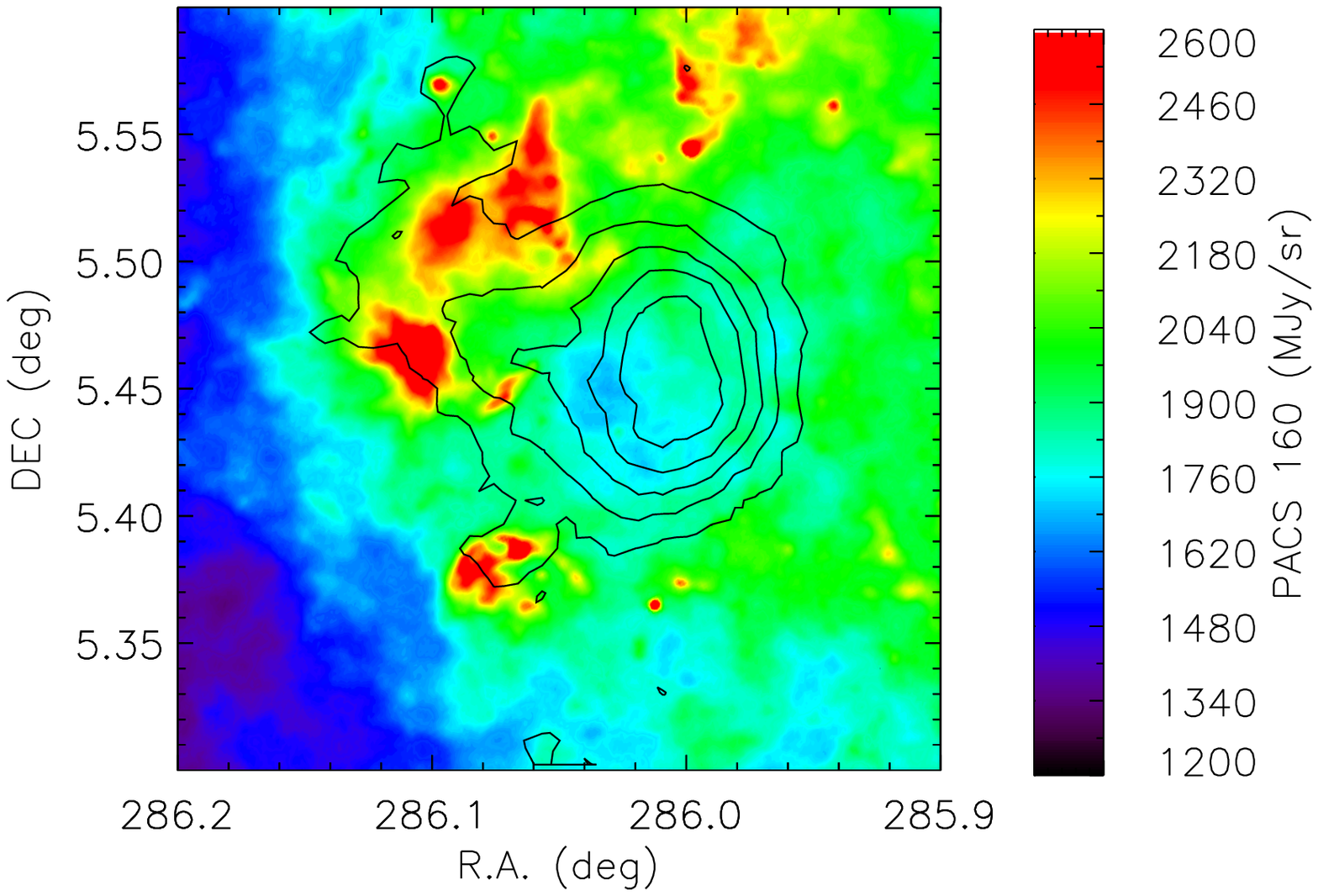}
\includegraphics[width=7cm]{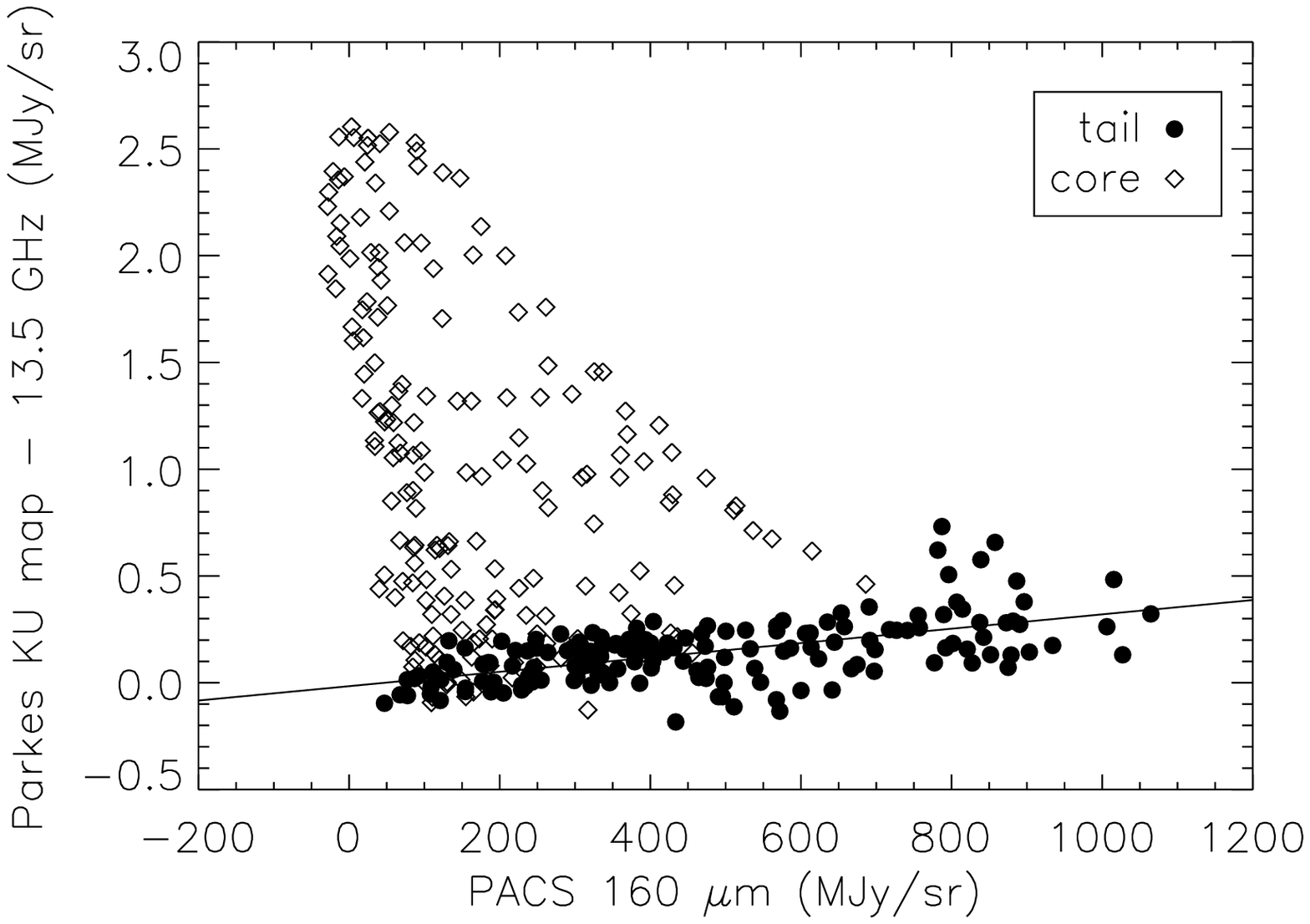}

\includegraphics[width=7cm]{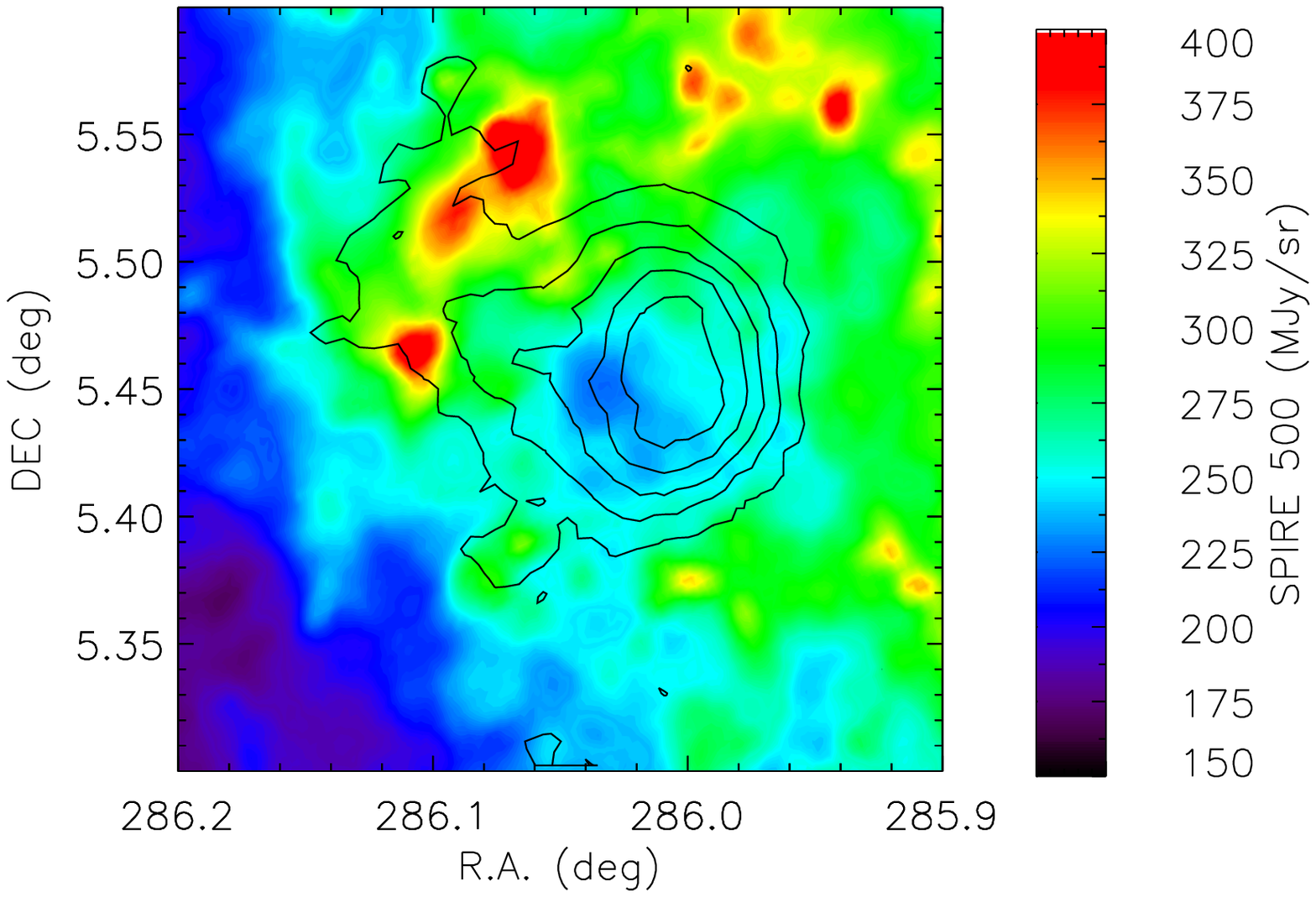}
\includegraphics[width=7cm]{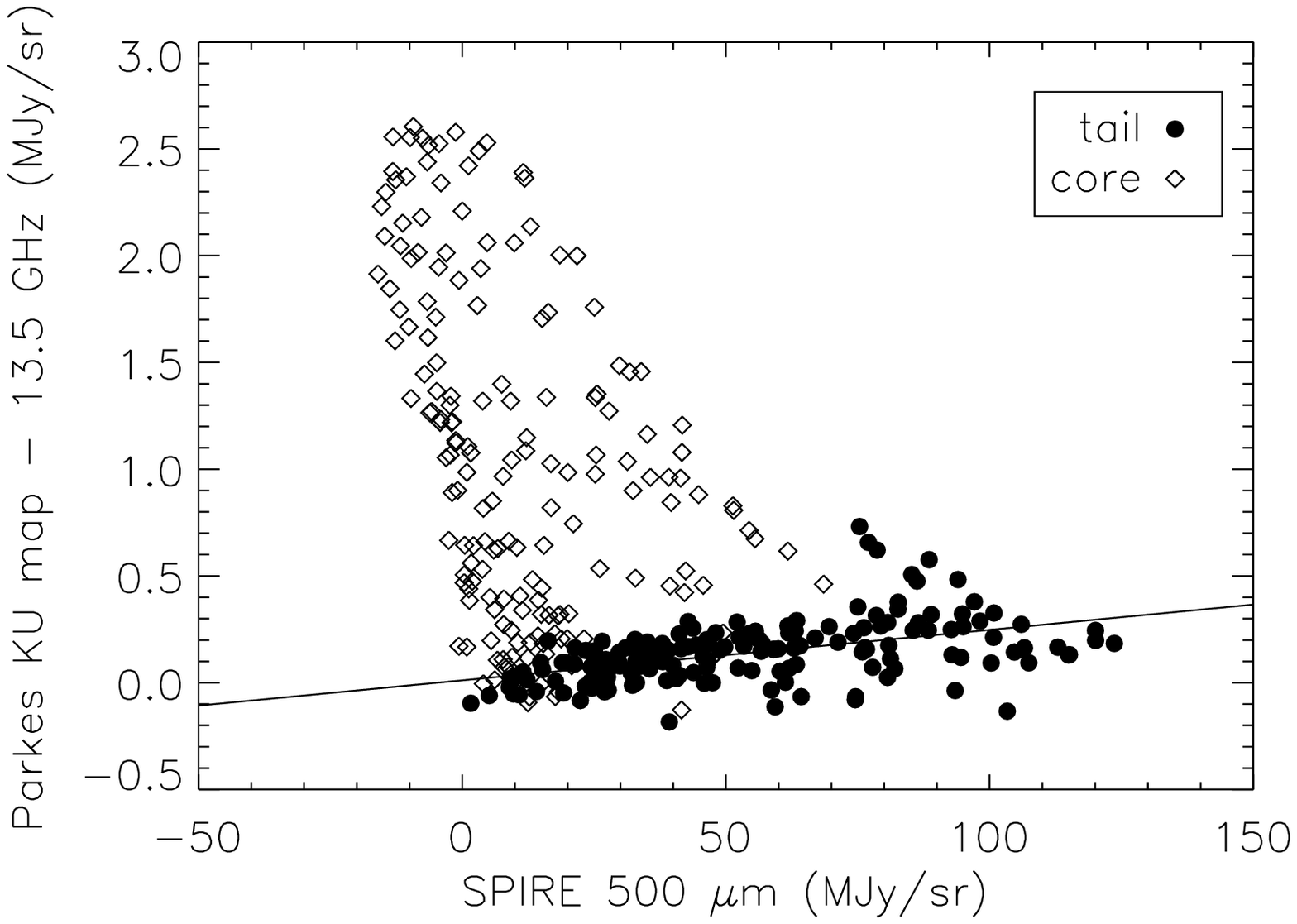}\label{tt}

\caption{Left column from top to bottom: 13.5~GHz Parkes map (contours) vs. Pacs 70~$\mu$m, Pacs 160~$\mu$m, 
and SPIRE 500~$\mu$m maps. The right column has the corresponding scatter plots, i.e., microwave map vs IR map.
 White diamonds denote emission from the core of 3C~396 while black circles denote that of the tail, with the definition of these two regions shown in Figure~\ref{regioni}. Contour levels, from the 13.5~{GHz} map, are at 0.2, 0.70, 1.2, 1.70, and 2.2 MJy/sr.}

\label{fig:tt_plots}
\end{center}
\end{figure*}

In order to compare our microwave measurements with IR emission, we make use of the SPIRE, 
PACS and MIPS IR maps from the \textit{Herschel} \citep{mol10} and \textit{SPITZER} \citep{spi} experiments,
 ranging from 500~$\mu$m to 24~$\mu$m. In Figure~\ref{fig:tt_plots} we compare the Parkes Ku 
13.5~GHz map with those extracted from the SPIRE instrument (500~$\mu$m) and the PACS instrument 
(160~$\mu$m and 70~$\mu$m) after smoothing them to the angular resolution of the  Parkes 13.5~GHz band. 
3C~396 has been divided into two regions (tail and core), as shown in Figure~\ref{regioni}. 
A lack of FIR emission is evident towards the core region of 3C~396, whereas there are bright FIR 
structures towards the northeast, where the tail identified by \cite{patnaik} and also hinted at in 
our Parkes data is located. These FIR structures have been studied in different publications (\cite{rea06}, \cite{lee09}, \cite{and11}).

In the three right panels of Figure~\ref{fig:tt_plots} we show TT correlation plots between the FIR and the radio emission at 13.5~GHz from Parkes.
A significant correlation is evident between the data sets for the tail region, whereas for the 
core region we see possible anti-correlation or no correlation at all. 

Pearson correlation coefficients for the tail emission are reported in Table~\ref{tbl_ir}. The 
statistical significance is at the 0.2\% level and we make significant measurements of all the coefficients. The slope 
of a linear fit to the pixels defining the tail-region gives an estimate of the emissivity in
the range 70--3500~$\mu$K/(MJy~sr$^{-1}$). These values are considerably higher than  what was found for
typical AME and HII regions. The correlation coefficient $R$ between 13.5~GHz Parkes map and the 500~$\mu$m map is lower than typical values estimated by the \textit{Planck}
experiment. Moreover, a close comparison between the IR maps and the microwave maps does not allow us to exclude the possibility that the IR emission is 
unassociated with the SNR and its microwave emission.

\begin{table}
\begin{center}
\caption{Results of the correlation analysis of tail emission between 13.5~GHz and IR experiments  }\label{tbl_ir}
\begin{tabular}{cccc}
\hline
Instrument  &  Pearson &   slope  & slope \\
						&         &               & ($\mu$k/(MJy/sr))   \\
\hline
SPITZER 24 &       0.369 &      $(1.15 \pm  0.23) \times 10^{-2}$  &  2050 $\pm$     410 \\
PACS 70 & 0.674 &       $(1.13 \pm 0.10) \times 10^{-3}$   & 202 $\pm$      17 \\
PACS 160 &       0.574 &      $ (3.36 \pm 0.38) \times 10^{-4}$  &   60 $\pm$      10 \\
SPIRE 250 &       0.537 &     $  (4.67 \pm  0.58) \times 10^{-4}$   &  83 $\pm$      11 \\
SPIRE 350 &       0.504 &     $  (9.9 \pm 1.4) \times 10^{-4}$   & 180 $\pm$      25 \\
SPIRE 500 &       0.480 &     $  (2.37 \pm   0.34) \times 10^{-3}$    & 420 $\pm$     60 \\
\hline
\end{tabular}
\end{center}
\end{table}

\section{Conclusions}\label{con}
We have performed new intensity observations of the SNR 3C~396 with the Parkes telescope at 8.4, 13.5, and
18.6~GHz and analysed them together with unpublished 31.2~GHz data from the GBT telescope. 
 We have calculated the SED of the core of the source and and its surroundings. Our observations argue against
 the presence of an AME component from the core of the source. When accounting for the surrounding 
 region, the SED is well-described by a power law with a Crab-like spectral index.
We find a spatial spectral index variation between 5~GHz and 31.2~GHz that is consistent with the variation reported by \cite{anderson} between 1.4~GHz and 5~GHz.

We have also performed new polarisation observations at 21.5~GHz that seem to confirm
 that 3C~396 emission is dominated by synchrotron, without any important contamination due to free--free as suggested by \cite{onic}.

The absence of correlation between IR and 13.5~GHz emission in the core of 3C~396 is a further finding, confirming that the emission of 3C~396 is not anomalous. On the other hand, we find a significant 
correlation in the tail-region at short wavelengths (i.e., $ < 160~\mu$m), as in other AME sources; however, this is probably due to diffuse emission from the Galactic plane.

\section*{Acknowledgments}

We acknowledge the logistic support provided by Parkes operators. The Parkes radio telescope is part of the Australia 
Telescope National Facility, which is funded by the Commonwealth of Australia for operation as a National Facility managed 
by CSIRO. The American National Radio Astronomy Observatory is a facility of the National Science Foundation
 operated under cooperative agreement by Associated Universities, Inc. We would like to thank A.D. Hincks, M. Hobson, A. Scaife for useful suggestions and comments. We thank the referee for providing constructive comments and help in improving the contents of this paper.

\label{lastpage}

\end{document}